\documentclass[journal]{IEEEtran}
\usepackage{cite}
\usepackage{amsmath,amssymb,amsfonts}
\usepackage{algorithmic}
\usepackage{graphicx}
\usepackage{textcomp}
\usepackage{todonotes}
\graphicspath{{./figs/}}
\begin{document}
\title{Dynamic Imaging using Deep Bi-linear Unsupervised Regularization (DEBLUR)}
\author{Abdul Haseeb Ahmed, \IEEEmembership{Member IEEE}, Prashant Nagpal,and Mathews Jacob, \IEEEmembership{Member IEEE}
\thanks{This work was supported in part by the NIH under Grant . }
\thanks{Abdul Haseeb Ahmed is with the Department of Electrical and Computer Engineering, The University of Iowa, Iowa City, IA 52242 USA (e-mail:abdul-ahmed@uiowa.edu).}
\thanks {Prashant Nagpal is with the Department of Radiology, The University of Iowa,
Iowa City, IA 52242 USA (e-mail: prashant-nagpal@uiowa.edu).}
\thanks{Mathews Jacob is with the Department of Electrical and Computer Engineering,
The University of Iowa, Iowa City, IA 52242 USA (e-mail: mathews-jacob@uiowa.edu).}}

\maketitle

\begin{abstract}
Bilinear models that decompose dynamic data to spatial and temporal factors are powerful and memory-efficient tools for the recovery of dynamic MRI data. These methods rely on sparsity and energy compaction priors on the factors to regularize the recovery. The quality of the recovered images depend on the specific priors. Motivated by deep image prior, we introduce a novel bilinear model whose factors are represented using convolutional neural networks (CNNs). The CNN parameters are learned from the undersampled data off the same subject. To reduce the run time and to improve performance, we initialize the CNN parameters. We use sparsity regularization of the network parameters to minimize the overfitting of the network to measurement noise. Our experiments on free-breathing and ungated cardiac cine data acquired using a navigated golden-angle gradient-echo radial sequence show the ability of our method to provide reduced spatial blurring as compared to low-rank and SToRM reconstructions.
\end{abstract}


\section{Introduction}
\label{sec:introduction}
\IEEEPARstart{D}{ynamic} MRI (DMRI) plays an important role in clinical applications such as cardiac cine MRI, which is commonly used by clinicians for the anatomical and functional assessment of organs. The clinical practice is to acquire the cine data using breath holding to achieve good spatial and temporal resolution. However, it is difficult for many subjects, including children, patients with myocardial infarction, and chronic obstructive pulmonary disease patients, to hold their breath \cite{copd}. In addition, multiple breath-holds prolong the scan time, adversely impacting patient comfort and compliance. 

Several computational approaches have been introduced to reduce the breath-held duration in cardiac cine and to enable free-breathing imaging. Early approaches relied on carefully designed signal models to exploit the structure of the data in x-f space \cite{dime,paradise,blast}, sparsity \cite{sparse}, or binned the data to different phases \cite{feng2016xd} which facilitates the signal recovery from undersampled measurements. In recent years, bilinear models\cite{zhao2012,ktfocuss,ktslr,ong2020extreme}, which represent the signal as the product of spatial and temporal factors, have emerged as powerful alternatives for the recovery of large scale data. These adaptive approaches, where the signal model is learned from the data itself, are observed to be far more efficient than earlier strategies that relied on hand-crafted models. The factor-based framework has been combined with other priors, including low-rank and sparsity \cite{ktslr,zhao2012}, low-rank +sparse \cite{otazo}, blind compressed sensing that learns a dictionary from the data \cite{bcs}, and motion-compensation \cite{master2013,lingala2014deformation}. Recently, non-linear manifold models that rely on kernel low-rank relation have been shown to outperform the subspace-based models in the context of free-breathing and ungated cardiac MRI \cite{sunritatci,leslieklr,ahmed2020}. Many of these schemes rely on k-space navigator measurements to estimate the temporal factors, while the spatial factors are estimated from the entire data. The kernel approach for estimating the temporal factors is observed to be more efficient in representing the dynamics, especially in free-breathing applications. A major benefit of the bilinear methods is the significantly reduced memory demand of these algorithms, in addition to the good reconstructions they offer. Specifically, the factors are significantly smaller in dimension than the dynamic dataset, which facilitates the recovery of large 3D volumes \cite{lustig-ong-lowrank,ong2020extreme}. While early methods relied on calibration data to estimate one of the factors, the joint optimization of both the factors offers several advantages, including improved image quality \cite{ktslr,bcs}. The distinction between the methods can be viewed as the specific priors applied on the spatial and temporal factors, including energy priors in the low-rank setting \cite{ktslr}, unit column norm and sparsity priors in the blind dictionary learning setting \cite{bcs}, and kernel priors in the manifold setting \cite{sunritatci,leslieklr}. 
\begin{figure*}[htb]
	\centering
	\includegraphics[ width=0.7\textwidth]{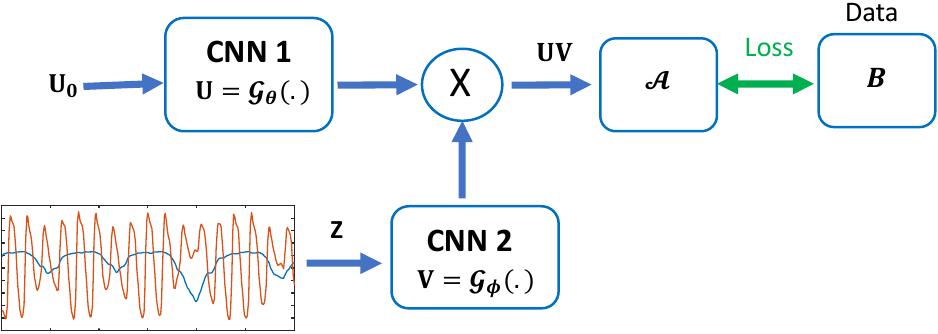}\\
	\caption {\small {Outline of the DEBLUR model. Two CNN generative networks $\mathcal G_{\theta}$ and $\mathcal G_{\phi}$ are used to generate the spatial ($\mathbf U$) and temporal $\mathbf V$ factors, respectively. The Casorati matrix of the dataset is modeled as $\mathbf X=\mathbf U\mathbf V^T$, where the columns of $\mathbf X$ correspond to the temporal frame. Rather than using random inputs as in \cite{dip}, the $\mathcal G_{\theta}$ generator is fed with the $\mathbf U$ factor matrix from SToRM, denoted by $\mathbf U_0$. The input to $\mathcal G_{\phi}$ network are the latent vectors $\mathbf Z$, which are also learned from the data. We expect the joint learning procedure to result in interpretable latent vectors, which will capture the temporal motion components (e.g., cardiac and respiratory motion).}}
	\label{fig1}
\end{figure*}

Deep learning models are now emerging as powerful approaches for image recovery in a range of static inverse problems \cite{wang2016}. Direct inversion strategies, which rely on a large convolutional neural network (CNN) to recover the images from undersampled data \cite{jong,unser}, as well as model-based deep learning methods \cite{hammernick,sch2017,modl} that interleave smaller CNN blocks with data-consistency enforcing optimization modules, have been introduced. By enforcing the data consistency, model-based methods can offer improved image quality over direct inversion strategies. Unfortunately, dynamic MRI and parametric MRI schemes often require the recovery of a large number of image frames; the direct application of the unrolled model-based deep learning schemes to the above setting is severely limited due to the high memory demand and computational complexity of current methods. Current strategies are either restricted to fewer time frames \cite{prieto} or often have to use small networks \cite{modlstorm,cheng}. Another challenge associated with these schemes is the lack of fully sampled training data for training these models, especially in the free-breathing and ungated mode.

The main focus of this work is to introduce a memory-efficient model for dynamic MRI called Deep Bi-Linear Unsupervised Representation (DEBLUR). This approach exploits the power of CNN to further improve the performance while retaining the memory efficiency of bilinear representations. We use the structural bias of CNN networks \cite{dip} as priors on the factors; this approach is an alternative to current schemes that penalize energy \cite{ktslr}, sparsity \cite{bcs}, or kernel priors \cite{modl,leslieklr} on the factors. In particular, we assume that the spatial and temporal factors are generated by two CNN generators whose parameters are estimated from the measured data. The CNN parameters are learned such that the k-space measurements of the bilinear representation matches the multi-channel measurements. The ability of the proposed scheme to directly learn the compressed representation makes it more memory efficient in multidimensional applications compared to current CNN approaches \cite{modlstorm, prieto, cheng}. 

This paper generalizes our earlier conference version \cite{ahmed2021dynamic} in multiple ways. First of all, we follow a completely unsupervised strategy; we pretrain the CNN factor generators from SToRM reconstruction of undersampled k-space data rather than using exemplar datasets. In addition, the generator of temporal basis functions is significantly different from \cite{ahmed2021dynamic}; the inputs to the generator are learned during the optimization in the current setting, which offers improved performance compared to keeping them fixed. More importantly, the approach is now  validated with several datasets compared to the limited comparisons in \cite{ahmed2021dynamic}, along with ablation studies to determine the impact of the regularization terms.

\section{Background}
\subsection{Bilinear models for dynamic MRI}
\label{bilinear_section}
Bilinear models are widely used in multidimensional applications, including dynamic MRI \cite{zhao2012,ktfocuss}, parameter mapping \cite{zhao2014model,bhave2016accelerated}, and MR spectroscopic imaging \cite{lam2014subspace,bhattacharya2017compartmentalized}. We have shown the significance of this approach in reconstructing dynamic images in our limited study \cite{ahmed2021dynamic}. These schemes express the Casorati matrix of the multidimensional dataset denoted by $\mathbf X=\left[\mathbf x_1,\ldots \mathbf x_N\right]$, where $\mathbf x_i$ denotes the vectorized version of the $i^{\rm th}$ frame in the dataset as 
\begin{equation}\label{bi-linear}
\mathbf X = \underbrace{\begin{bmatrix}
\mathbf u_1,\ldots, \mathbf u_r\end{bmatrix}}_{
\mathbf U} \underbrace{\begin{bmatrix}
\mathbf v_1,\ldots, \mathbf v_r\end{bmatrix}^T}_{
\mathbf V^T}.
\end{equation}
The columns of $\mathbf U$ denoted by $\mathbf u_i$ are identified as the spatial basis functions, while those of $\mathbf V$ are the temporal basis functions. The main benefit in the above representation is the significant reduction in the number of free parameters that need to be estimated, which translates to reduced data demand \cite{modlstorm}. In the context of large multidimensional applications, another key benefit is the memory and computational benefits. In particular, the measurements can be expressed completely in terms of $\mathbf U$ and $\mathbf V$ as:
\begin{equation}\label{key}
\mathcal A_i(\mathbf x_i) =\mathbf A_i(\mathbf U) \mathbf v_i
\end{equation}
This approach eliminates the need for computing and storing the image frames $\mathbf x_i$ themselves during image reconstruction; post-recovery, the desired frame can be retrieved as $\mathbf x_i = \mathbf U \mathbf v_i$. 

Many schemes \cite{zhao2012,ktfocuss} estimate the temporal basis functions $\mathbf V$ from k-space navigators, followed by the estimation of the spatial factors $\mathbf U$ in \eqref{bi-linear}. By contrast, the joint estimation of $\mathbf U$ and $\mathbf V$ from the measured data offers improved performance and reduces the need for specialized acquisition schemes with k-space navigators. The joint approaches pose the recovery of the signals from the undersampled measurements $\mathcal A(\mathbf X)$ as:
\begin{equation}\label{bisolv}
\{\mathbf U^*,\mathbf V^*\} = \arg \min_{\mathbf U,\mathbf V} \left\|\mathcal A(\mathbf U\mathbf V^T) -\mathbf B\right\|^2 + \lambda_1 \mathcal R_1(\mathbf U) + \lambda_2 \mathcal R_2 (\mathbf V)
\end{equation}
Here, $\mathbf B$ denotes the measured data. We note that the representation in \eqref{bi-linear} is linear in $\mathbf U$ and $\mathbf V$ independently; the joint optimization in \eqref{bisolv} can be viewed as a bilinear optimization. Here, $\mathcal R_1$ and $\mathcal R_2$ are regularization functionals. Depending on the specific form of the regularization functions, one would obtain different flavors of reconstruction algorithms.
\begin{enumerate}
	\item  Low-rank regularization \cite{ktslr}: Here, one would choose $\mathcal R_1(\mathbf U) = \|\mathbf U\|^2$ and $\mathcal R_2(\mathbf V) = \|\mathbf V\|^2$ .
	\item  Blind compressed sensing: Here, one would choose $\mathcal R_1(\mathbf U) = \|\mathbf U\|_{\ell_1}$ and $\mathcal R_2(\mathbf V) = \|\mathbf V\|^2$ .
	
	\item Smoothness regularization on manifolds (SToRM): The SToRM scheme also relies on a factorization as in \eqref{bi-linear}, where $R_1 (\mathbf U) = \sum \sigma_i \|\mathbf u\|_i^2$ and $\mathbf V$ is obtained as the eigenvectors of the graph Laplacian matrix of the graph of the data. Both calibrated \cite{SToRM} and uncalibrated  \cite{ahmed2020} formulations are available.
\end{enumerate}
The performance of the above methods critically depends on the specific choice of the priors  $\mathcal R_1$ and $\mathcal R_2$ to estimate $\mathbf U$ and $\mathbf V$. 

\vspace{-1em}
\subsection{Deep Image Prior (DIP)}
The deep image prior approach has been introduced in inverse problems to exploit the structural bias of CNNs to natural image structure rather than noise \cite{dip}. The regularized reconstruction of a static image from undersampled and noisy measurements are posed as 
\begin{equation}\label{dip}
\{\boldsymbol \theta^*\} = \arg \min_{\theta}\left\|\mathcal A(\mathbf x) - \mathbf b\right\|^2 ~~\mbox{such that} ~~ \mathbf x = \mathcal G_{\boldsymbol \theta}[\mathbf z]
\end{equation} 
where $\mathbf x=\mathcal G_{\boldsymbol \theta^*}(\mathbf z)$ is the recovered image, generated by the CNN generator $\mathcal G_{\boldsymbol\theta^*}$ whose parameters are denoted by $\boldsymbol\theta$. 

The constraint that the image is generated by a CNN provides implicit regularization, which facilitates the recovery of $\mathbf x$ in challenging inverse problems. Here, $\mathbf z^*$ is a random latent variable, which may or may not be optimized. The structural bias of untrained CNN generators towards natural images is exploited to yield good recovery. The above problem is often solved using stochastic gradient descent (SGD), which is often terminated early to obtain regularized recovery. Few iterations of the above model are observed to represent images reasonably well, while the model will learn the noise in the measurements, resulting in poor reconstructions with more iterations, provided the generator has sufficient capacity. Early termination is often used to avoid this and thus regularize the recovery. Alternate approaches including alternatives to SGD have been introduced to avoid the early stopping strategies. 

\begin{table}
	\caption{Quantitative comparison of the DEBLUR method with SOTA methods.}
	\centering
	\setlength{\tabcolsep}{3pt}
	\begin{tabular}{|p{25pt}|p{45pt}|p{45pt}|p{45pt}|}
	    \hline
		Metric & Low-Rank & SToRM(14s) & DEBLUR(14s)\\
		\hline
		SER & 
		$18.93\pm0.48$ & 
		$22.41\pm0.78$ &
		$35.99\pm4.98$\\
		\hline
		SSIM & 
		$0.38\pm0.02$ & 
		$0.59\pm0.04$ &
		$0.96\pm0.04$\\
		\hline
		HFEN & 
		$1.23\pm0.02$ & 
		$0.82\pm0.05$ &
		$0.18\pm0.13$\\
		\hline
		Brisque & 
		$40.82\pm2.1$ & 
		$32.88\pm4.3$ &
		$26.83\pm5.81$\\
		\hline
	\end{tabular}
	\label{tab2}
\end{table}

\begin{figure}[htb]
	\centering
	\includegraphics[ scale=1.1]{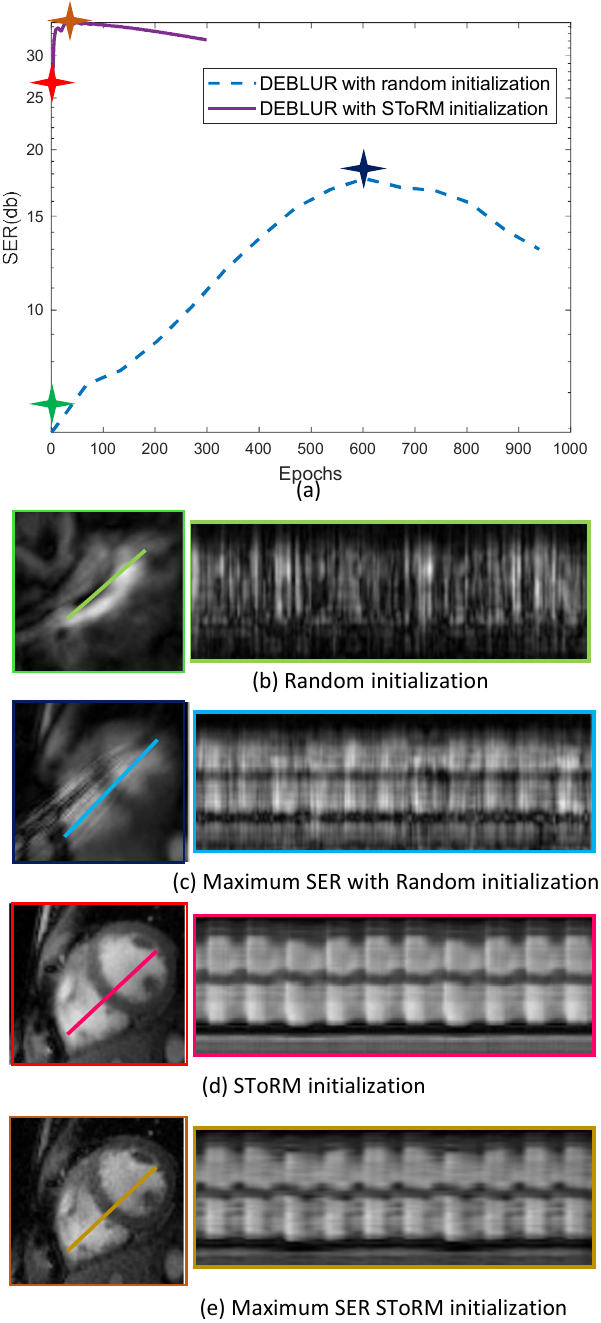}\\
	\caption {\small {(a) shows the SER curves of the DEBLUR scheme with random initialization and with SToRM initialization. Their corresponding initial and peak values images are shown in (b)-(e). The image with the green border (b) corresponds to the initialization with random weights, while the solution with the peak SER with random initialization is shown in (c). The use of the pretrained parameters yields (d), while optimizing the parameters significantly improves the performance as seen from (e). We note the reduction in spatial and temporal blurring and the absence of artifacts and sharper features in (e). }}
	\label{fig2}
\end{figure}

\begin{figure}[htb]
	\centering
	\includegraphics[ scale=0.7]{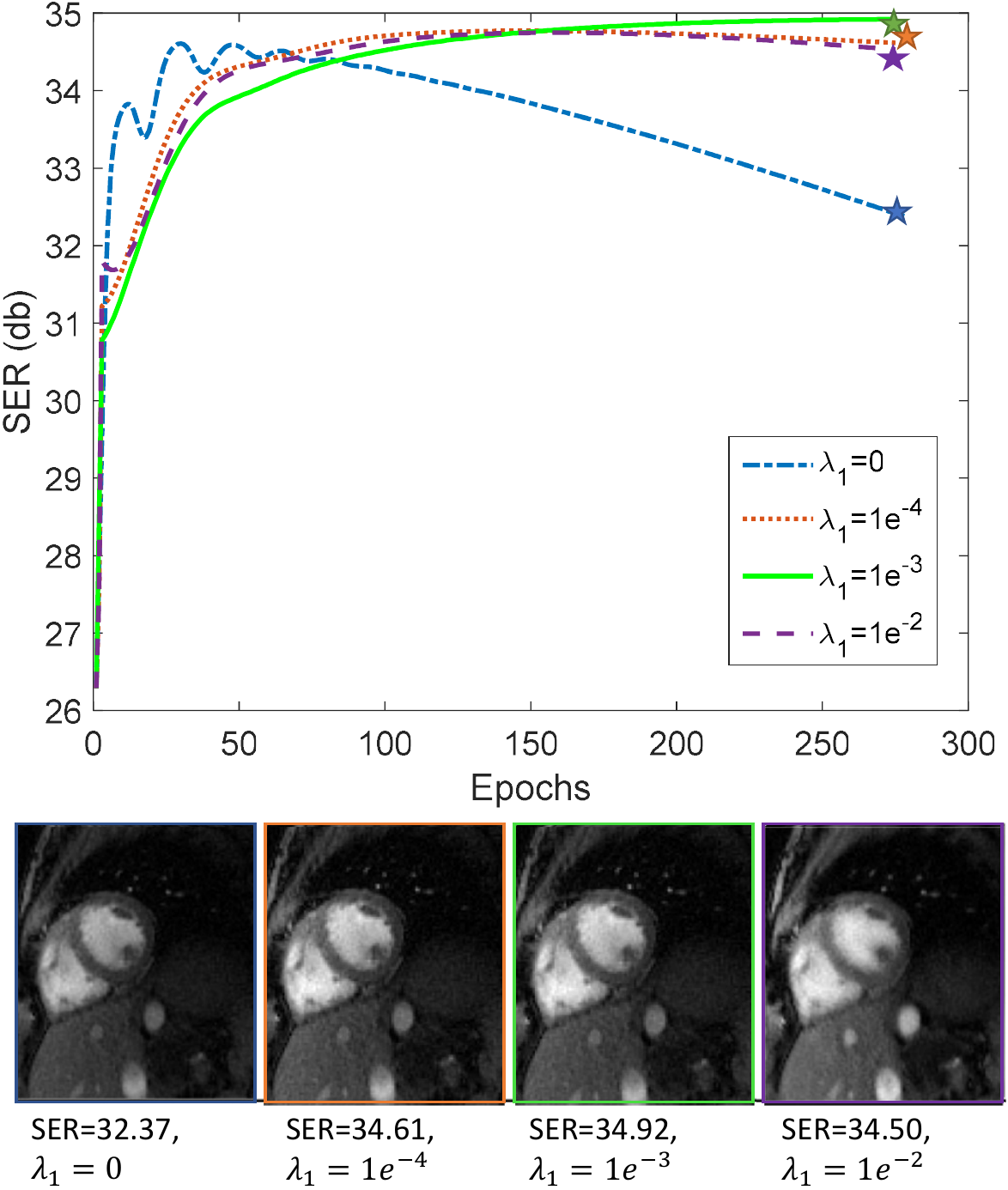}\\
	\caption {\small { shows the SER curves with different $\lambda_1$ values in DEBLUR method in \eqref{DEBLURrg}, which regularizes the generator $\mathcal G_{\theta}$ used to generate $\mathbf U$. The un-regularized setting is denoted by the blue curve, which is the zoomed version of the red-curve in Fig. \ref{fig2}. We note that higher regularization parameters control the decay of SER with iterations, which indicates improved generalization of the network to unseen k-space samples. We note that the best performance is achieved with $\lambda_1=10^{-3}$. Larger regularization parameters (e.g., $\lambda=10^{-2}$ denoted by the red curve) translate to slight oversmoothing of the spatial factors.  }}
	\label{fig3}
\end{figure}

\begin{figure}[htb]
	\centering
	\includegraphics[ scale=0.7]{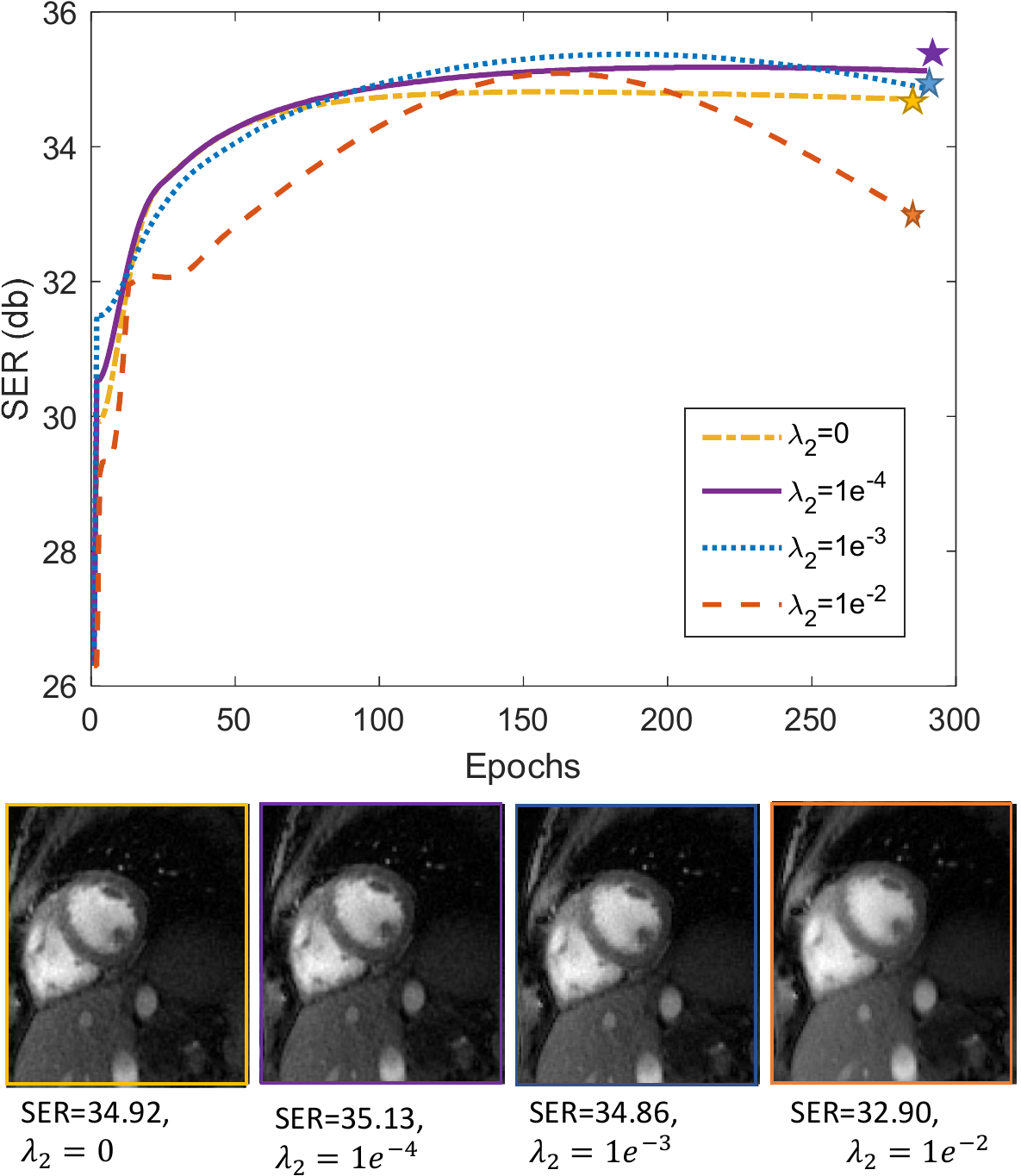}\\
	\caption {\small { shows the SER curves with different $\lambda_2$ values in DEBLUR method. $\lambda_2$ regularizes $\mathbf V$ network parameters. Lower $\lambda_2$ value allows network to learn noisy temporal information in the data as illustrated by the yellow curve. Higher value of $\lambda_2$ oversmooths the temporal basis as depicted by the orange color. Empirical findings show $\lambda_2$ $=1e^{-4}$ gives better image SER as compared to higher and lower values. Secondly, the performance does not deteriorate if it runs for more epochs, as shown by the purple color. }}
	\label{fig4}
\end{figure}
\begin{figure}[htb]
	\centering
	\includegraphics[ scale=0.75]{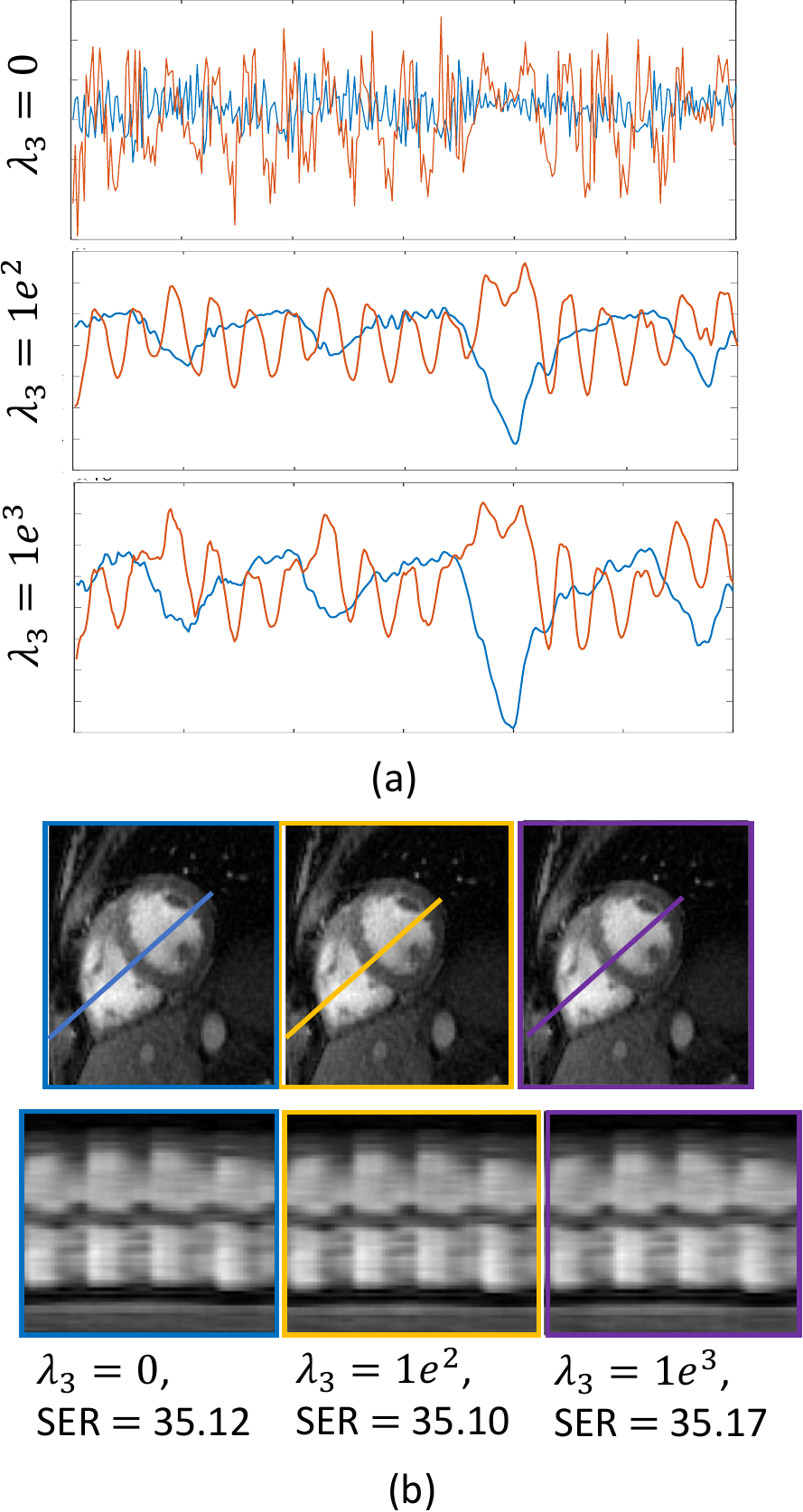}\\
	\caption {\small { shows the latent vectors with different $\lambda_3$ values in the DEBLUR method. $\lambda_3$ applies temporal smoothness on latent vectors to achieve meaningful cardiac and respiratory motion. Lower $\lambda_3$ gives noisy latent vectors as illustrated by $\lambda_3=0$. Higher value of $\lambda_3$ oversmooths the latent vectors as shown by $\lambda_3=1e^3$. Empirical findings show $\lambda_1$ $=1e^2$ gives better separation of latent vectors that represent cardiac and respiratory motion. Secondly, the SER changes between the different values of $\lambda_3$ are marginal. }}
	\label{fig4a}
\end{figure}
\section{Deep Bi-Linear Representation (DEBLUR)}

The main focus of this work is to develop a multidimensional image reconstruction algorithm that inherits the memory efficiency of bilinear models; we express the data as $\mathbf X = \mathbf U\mathbf V^T$ as in \eqref{bi-linear}. Unlike current bilinear models that use energy and sparsity priors, we use CNN priors on the factors. In particular, we propose to represent the factors as 
\begin{eqnarray}\label{unet}
\mathbf U &=& \mathcal G_{\theta}(\mathbf U_0)\\\label{vnet}
\mathbf V &=& \mathcal G_{\phi}(\mathbf Z).
\end{eqnarray}
Here, $\mathcal G_{\theta}$ and $\mathcal G_{\phi}$ are CNN generators whose weights are denoted by $\theta$ and $\phi$, respectively. Here, $\mathbf Z = \left[\mathbf z_1,..,\mathbf z_N\right]$ are time-dependent latent vectors that are also learned from the measured k-space data. Here, $\mathbf U_0$ is an initial approximate reconstruction that is often obtained using a simple algorithm, which is refined by the generator.  See Fig. \ref{fig1} for details. We observe that feeding an initial reconstruction of the $\mathbf U$ factors to $G_{\theta}$ offers faster reconstructions compared to the DIP strategy of feeding noise to the image generator.  Similar to DIP \cite{dip}, we expect to capitalize on the structural bias of the CNNs towards smooth natural images.

We propose to recover the factors by solving the optimization problem:
\begin{eqnarray}\nonumber\label{DEBLUR}
\{\mathbf U^*,\mathbf V^*\} &=& \min_{\theta,\phi,\mathbf Z} \left\|\mathcal A(\mathbf U\mathbf V^T) -\mathbf B\right\|^2,\\\nonumber&&\qquad \mbox{ such that } \mathbf U = \mathcal G_{\theta}\left[\mathbf U_0\right]; \mathbf V = \mathcal G_{\phi}\left[\mathbf Z\right],\\
\end{eqnarray}
which is solved using SGD. Note that we also optimize for the latent vectors $\mathbf Z$. 

Note that the recovery scheme in \eqref{DEBLUR} recovers the factors directly from the multi-channel measurements $\mathbf B$. Once the factors are recovered, we can generate the $i^{\rm th}$ frame of the time series as
\begin{equation}\label{frame}
    \mathbf x_i = \mathbf U \underbrace{\mathcal G_{\phi}[\mathbf z_i]}
\end{equation}
This approach significantly reduces the memory demand and computational complexity of the algorithm, especially in applications involving multidimensional time series.

%\begin{figure}[htb]
%	\centering
%	\includegraphics[ scale=0.6]{f1-crop.pdf}\\
%	\caption {\small {Proposed method. Two CNN networks are used on the spatial and temporal prior factors. CNN 1 is initialed with the $\mathbf U_0$ as mentioned in Eq. 6. CNN 2 is initialized with the SToRM temporal basis. We have also applied $l_1$ norm on the network parameters to stabilize the convergence.}}
%	\label{fig1}
%\end{figure}

%\begin{figure}[htb]
%	\centering
%	\includegraphics[ scale=0.4]{f3.png}\\
%	\caption {\small {shows the benefits of using $l_1$ regularization on network parameters. The dotted line shows the plot of the DEBLUR method without any regularization. Other curves show $l_1$ regularization on the $U$ CNN network, the $V$ CNN network, and both.Use of $l_1$ regularization provides the benefits of improved signal-to-noise ratio (SNR) and stable convergence.  }\textcolor{red}{Add more details. Change legend to $\lambda_1=0$ etc. No significant benefit with adding U regularization with V. Merge Fig 2 and Fig 3 to save space ?? }}
%	\label{fig3}
%\end{figure}\vspace{-1em}

\subsection{Unsupervised pretraining of the generators}
The generators in DIP are usually initialized with random weights. Unlike past convex strategies, the CNN-based algorithm in \eqref{DEBLUR} is not guaranteed to converge to the global minimum of the cost function. The final solution will be dependent on the initialization. To improve performance and to reduce the computational complexity, we propose to initialize the $\mathcal G_{\theta}$ and $\mathcal G_{\phi}$ networks. In this work, we chose the SToRM \cite{sunritatci} data to initialize the network. We note from our past results that the SToRM approach yields improved results compared to the other state-of-the art algorithms  \cite{sunritatci}, including low-rank methods. 
In particular, we pretrain  $\mathcal G_{\theta}$ and $\mathcal G_{\phi}$ generators independently using the SToRM factors, as 
\begin{eqnarray}\label{thetainit}
\theta_0 &=& \arg \min_{\theta}\|\mathcal G_{\theta}(\mathbf U_{0})-\mathbf U_{\rm SToRM}\|^2,
\\\label{phiinit}
\{\phi_0,\mathbf Z_0 \}&=& \arg \min_{\phi,\mathbf Z}\|\mathcal G_{\phi}(\mathbf Z)-\mathbf V_{\rm SToRM}\|^2.
\end{eqnarray}
These initial guesses of the network weights $(\theta,\phi)$ and latent vectors $\mathbf Z$ are used as initialization in \eqref{DEBLUR}.  We study the impact of the initialization on the algorithms in the results section.

\subsection{Regularization penalties}
The DIP approach, as well as our extension to the dynamic setting in \eqref{DEBLUR}, is vulnerable to noise overfitting. In particular, if the generator networks have sufficient capacity, it will learn the noise in the measurements when the number of iterations is large \cite{dip}. To further improve the robustness of the algorithm to noise, we propose to additionally penalize the $\ell_1$ norm of the network weights. The regularized recovery is posed as 
\begin{eqnarray}\nonumber\label{DEBLURrg}
\{\mathbf U^*,\mathbf V^*\} &=& \min_{\theta,\phi,\mathbf Z} \left\|\mathcal A(\mathbf U\mathbf V^T) -\mathbf B\right\|^2\\\nonumber
&&\qquad + \underbrace{\lambda_1 \|\theta\|_{\ell_1} + \lambda_2 \|\phi\|_{\ell_1}}_{\mbox{\scriptsize network regularization}} + \underbrace{\lambda_3 \|\nabla_t \mathbf{Z}\|_{\ell_1}}_{\mbox{\scriptsize temporal regularization}},\\\nonumber\\\nonumber&&\qquad \mbox{ such that } \mathbf U = \mathcal G_{\theta}\left[\mathbf U_0\right]; \mathbf V = \mathcal G_{\phi}\left[\mathbf Z\right]\\
\end{eqnarray}
 We expect the weight-regularization strategy to learn networks with fewer non-zero weights, which translates to smaller capacity; this approach is expected to reduce the vulnerability of the network to overfitting. The sparsity of the weights of $\mathcal G_{\theta}$ promotes the learning of local spatial factors similar to \cite{ong2020extreme}, which is more efficient than global low-rank minimization. Unlike the explicit definition of the blocks in \cite{ong2020extreme}, the definition of the local neighborhoods in the proposed scheme are more data-driven. The images in the time series often vary smoothly with time in dynamic imaging applications; the norm of the temporal derivatives of the images is a widely used prior. Our goal is to directly recover the compressed representation from the data without using the actual images. We note from \eqref{frame} that the $i^{\rm th}$ image in the time series is dependent on the latent vector $\mathbf z_i$. 
The last term in \eqref{DEBLURrg} is the norm of the temporal gradients of $\mathbf Z$, which will encourage the temporal smoothness of the time series. The algorithm in \eqref{DEBLURrg} is also initialized with the pretraining strategy discussed in the previous section. The impact of the regularization parameters and their ability to minimize overfitting issues are studied in the results section.

% \begin{table}
% 	\caption{Detail of the datasets}
% 	\label{table}
% 	\centering
% 	\setlength{\tabcolsep}{3pt}
% 	\begin{tabular}{|p{25pt}|p{25pt}|p{90pt}|}
% 		\hline
% 		Name& 
% 		Sex& 
% 		Indication\\
% 		\hline
% 		Data1& 
% 		Female& 
% 		cardiomegaly\\
% 		Data2& 
% 		Female& 
% 		RV enlargement\\
% 		Data3& 
% 		Male& 
% 		Ischemic cardiomyopathy \\
% 		Data4& 
% 		Male& 
% 		atrial fibrillation \\
% 		Data5& 
% 		Female& 
% 		RVOT mass \\
% 		\hline
% 	\end{tabular}
% 	\label{tab1}
% \end{table}
\section{Implementation details}
\subsection{Data acquisition and post-processing}
The experimental data was obtained using the FLASH sequence on a Siemens 1.5T scanner with 34 coil elements total (body and spine coil arrays) in the free-breathing and ungated mode from cardiac MRI patients with a scan time of 42 seconds per slice.The patients with cardiac abnormalities were recruited from those who were referred for routine clinical examinations. The protocol was approved by the Institutional Review Board (IRB) at the University of Iowa.

\subsubsection{Pulse sequence details}
The datasets were acquired using a 34-channel cardiac array. We used a radial GRE sequence with the following parameters: TR/TE 4.68/2.1 ms, FOV 300 mm, base resolution 256, slice thickness 8 mm. A temporal resolution of 46.8 ms was obtained by sampling 10 k-space spokes per frame. Each temporal frame was sampled by two k-space navigator spokes (out of 10 spokes/frame), oriented at 0 degrees and 90 degrees, respectively. The remaining spokes were chosen with a golden-angle view ordering. The scan parameters were kept the same across all patients. The subjects were asked to breathe freely, and the data was acquired in an ungated fashion. The complete data acquisition lasted 42 seconds for each slice. To determine the ability of the algorithms to reduce the acquisition time, we retained the initial $14$ seconds of the original acquisition. 

\subsubsection{Coil selection and compression:}  To improve the image reconstruction quality, we excluded the coils with low sensitivities in the region/slice of interest. We used an automatic coil selection algorithm to pick the five best coil images, which provided the best signal-to-noise ratio (SNR) in the heart region. Our experiments (not included in this paper) show that this coil combination has minimal impact on image quality. The main motivation for the combination was to reduce the memory requirement so that it fit on our GPU device, which significantly reduced the computational complexity. All the results were generated using a single node of the high-performance Argon Cluster at the University of Iowa, equipped with Titan V 32GB of memory.  

\subsubsection{Performance Metrics:} We used the following four quantitative metrics to compare our method against the existing schemes: 
\begin{itemize}
    \item Signal to error ratio (SER):
    \begin{equation}
	\mathbf{SER} = 20\log_{10} {\frac{ \|\mathbf{x}_{\textrm{ orig}}\|_2}{||\mathbf{x}_{\textrm {orig}}-\mathbf{x}_{\textrm{rec}}||_{2}}},
	\end{equation}
	where $||\cdot||_{2}$ donates the $\ell_{2}$ norm, and $\mathbf{x}_{orig}$  and $\mathbf{x}_{rec}$ denote the original and the reconstructed images, respectively.
	\item Peak Signal to Noise Ratio (PSNR):
	\begin{equation}
	\mathbf{PSNR} = 20\log_{10} {\frac{\max \{\mathbf{x}_{\textrm{ orig}}\}}{||\mathbf{x}_{\textrm {orig}}-\mathbf{x}_{\textrm{rec}}||_{2}}}.
	\end{equation}
	\item Normalized High Frequency Error (HFEN) \cite{ravishankar2011}: This measures the quality of fine features, edges, and spatial blurring in the images and is defined as: 
	\begin{equation}
	\mathbf{HFEN} = 20\log_{10} {\frac{||\textrm{LoG}(\mathbf{x}_{\textrm {orig}})-\textrm{LoG}(\mathbf{x}_{\textrm {rec} })||_{2}}{||\textrm{LoG}(\mathbf{x}_{\textrm {orig}})||_{2}} },
	\end{equation}
	where $\textrm{LoG}$ is a Laplacian of the Gaussian filter that captures edges. We use the same filter specifications as did Ravishankar et al. \cite{ravishankar2011}: kernel size of 15 $\times$ 15 pixels, with a standard deviation of 1.5.
	\item The Structural Similarity index (\textbf{SSIM}) is a perceptual metric introduced in \cite{wang2004image}, whose implementation is publicly available. We used the default contrast values, Gaussian kernel size of 11 $\times$ 11 pixels with a standard deviation of 1.5 pixels.
	\item BRISQUE is a referenceless measure of image quality, where a smaller score indicates better perceptual quality. BRISQUE estimates and gets the score using a support vector regression (SVR) model with the help of an image database and corresponding differential mean opinion score values. The distorted image database such as compression artifacts, blurring, and noise images, and with pristine versions of the distorted images \cite{brisq}.
\end{itemize}
Higher values  of the above-mentioned performance metrics correspond to better reconstruction, except for the HFEN, where a lower value is better.\\
\subsubsection{State-of-the-art algorithms for comparison:}
We compare the proposed scheme against the following algorithms:
\begin{itemize}
	\item SToRM \cite{sunritatci}: The manifold Laplacian is estimated from the self-gating navigators acquired in k-space. Once the Laplacian matrix is obtained from navigators, the high-resolution images are recovered using kernel low-rank-based framework.
	\item Low-Rank \cite{zhao2012,ktslr}: The image time series is recovered by nuclear norm minimization. The nuclear norm minimization approach models the images as points living on a subspace. 
\end{itemize}

\begin{figure*}[htb]
	\centering
	\includegraphics[ scale=0.8]{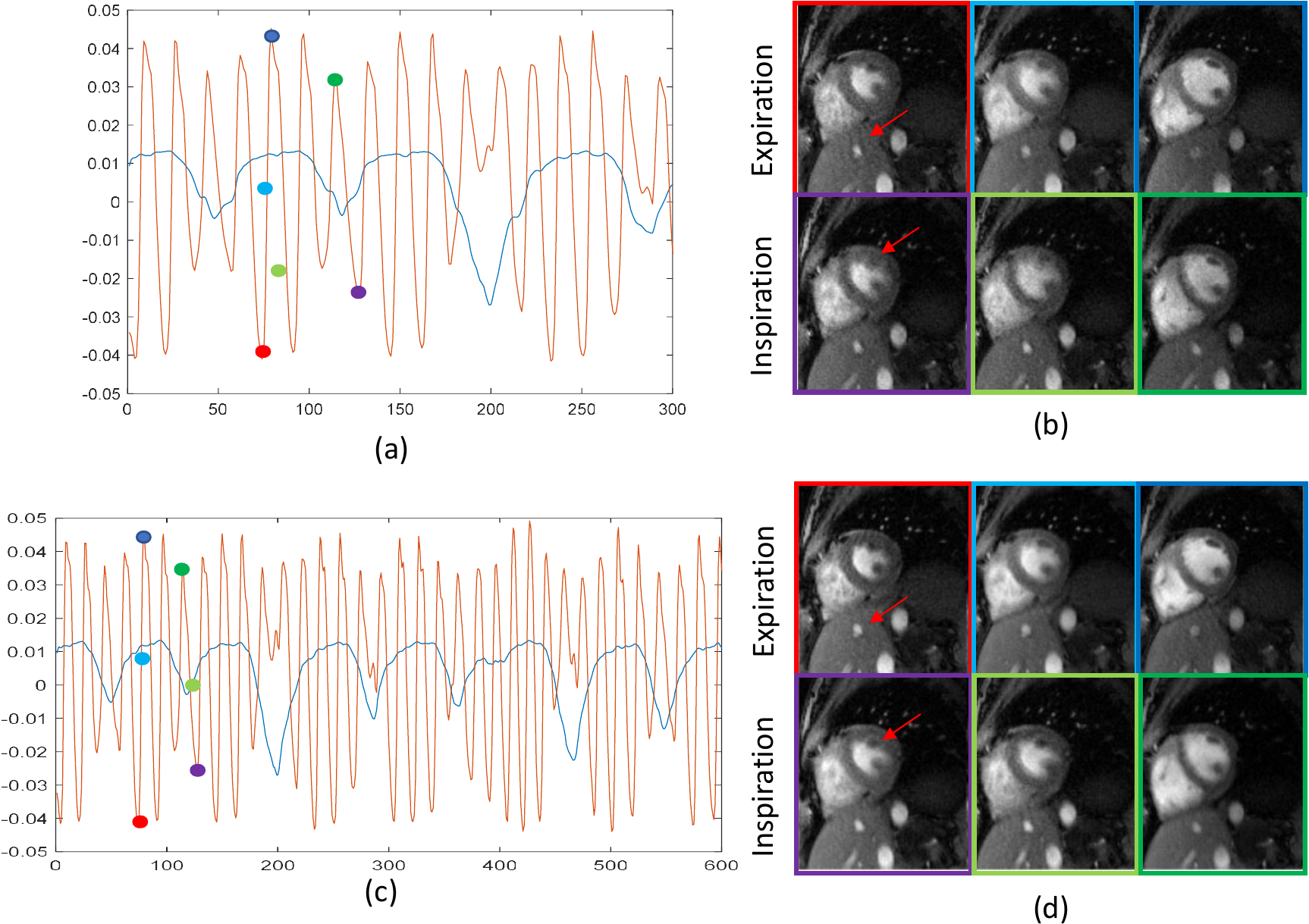}\\
	\caption {\small {Comparison of the DEBLUR method using 14 s and 28 s of acquired data. Fig \ref{fig5}(a) shows the two latent vectors of 14 s data, where the blue vector represents respiratory motion and the red vector represents cardiac motion in the data. Images at different time points, indicated by color dots, are shown in Fig \ref{fig5}(b). It also shows the ability of our latent vector-based approach to capture images at different time instants in a series. For example, the red dot in Fig \ref{fig5}(a) captures the image at the systole phase and at the expiration state. We have also compared 14 s data with 28 s data to show the improvement in the performance of the DEBLUR method. Differences in image quality with less and more data are subtle in the diastole phase but become prominent in the systole phase, as indicated by the red arrows.}}
	\label{fig5}
\end{figure*}

\begin{figure}[htb]
	\centering
	\includegraphics[ scale=0.7]{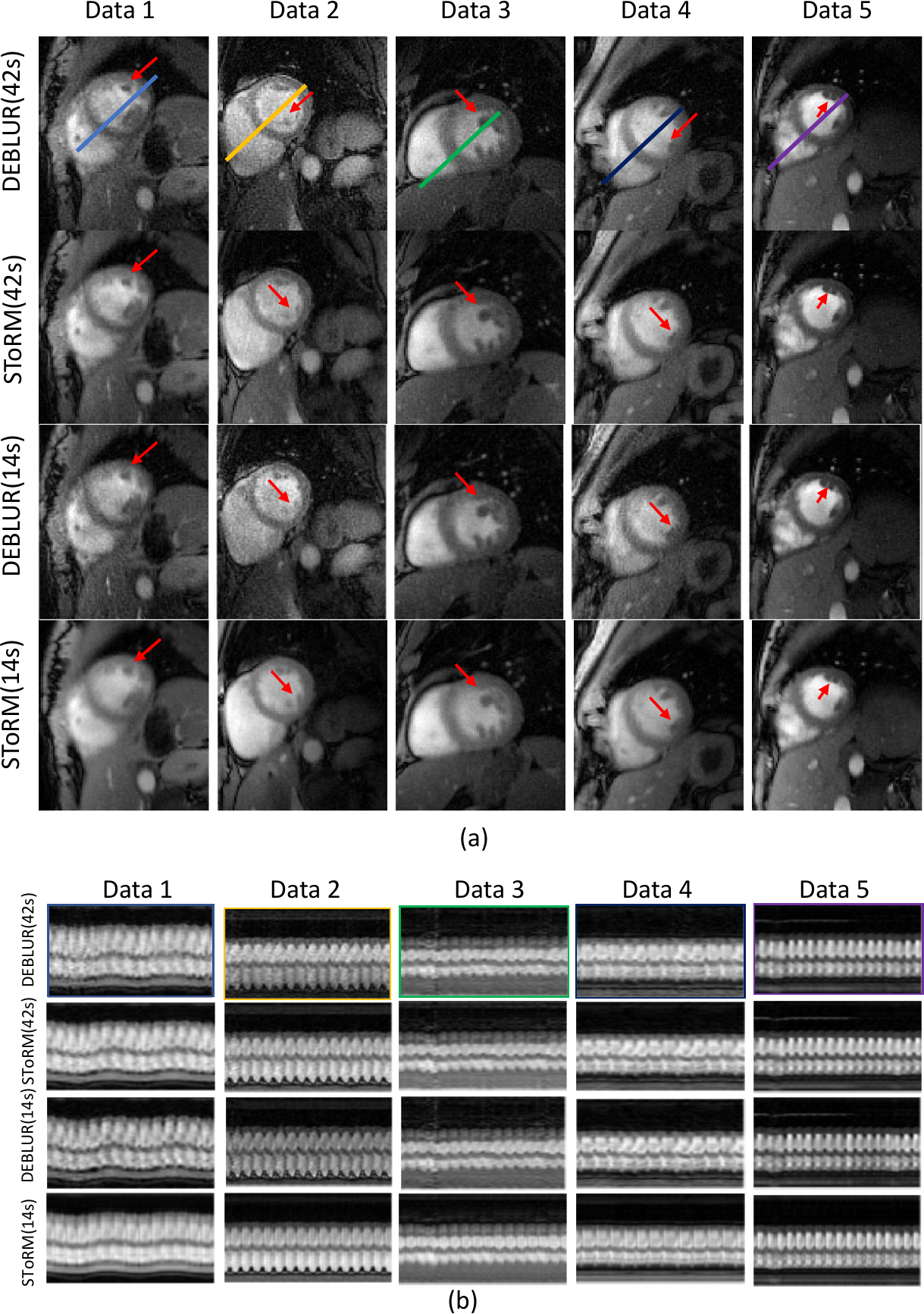}\\
	\caption {\small {Comparison of the DEBLUR and SToRM methods using 14 s and 42 s of acquired data. We have used five different datasets to compare the performance of the DEBLUR and SToRM methods. With less data (14 s), images have less sharpness than they do with 42 s of data. However, DEBLUR(14s) gives better image quality than the SToRM(14s) method. DEBLUR(42s) achieves better image contrast with enhanced features, as depicted by red arrows in the Figure.}}
	\label{fig6}
\end{figure}

\begin{figure*}[htb!]
	\centering
	\includegraphics[ scale=0.7]{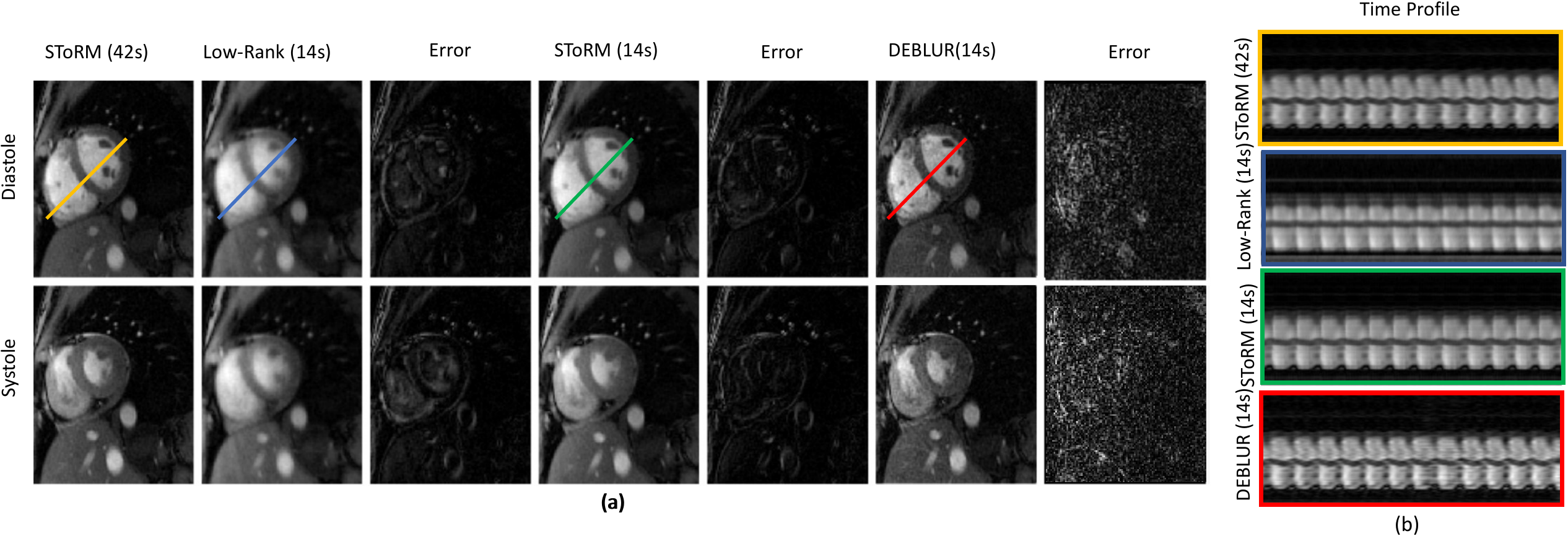}\\
	\caption {\small {Comparison of the DEBLUR(14s) method with the existing method. Since ground truth is not available, we have SToRM(42s) as ground truth. (a) To compare the image quality spatially, we have shown two frames (diastole and systole) from each method, along with their error maps. (b) Time profiles are shown.}}
	\label{fig7}
\end{figure*}
\subsection{Architecture of the generators}
We refer to $\mathcal G_{\theta}$ as the spatial generator. We use a four-layer CNN network with ReLU activation. The number of channels of the input and output is equal to twice the number of basis functions, to account for the real and imaginary parts of the basis. In our work, we use 30 basis functions, and hence the number of input and output channels is 60. We refer to $\mathcal G_{\phi}$ as the temporal generator, where we use a four-layer CNN network with ReLU activation. The inputs to the temporal generator are the latent vectors $\mathbb Z \in$ $\Re ^ {d \times nf}$, where $d$ represents the latent dimension and $n_f$ denotes the number of frames in the dynamic dataset. In our work, we observe that two-dimensional latent vectors are sufficient to obtain good reconstructions. The outputs of the temporal generator are the temporal basis functions $\mathbf V$ of dimension $r\times n_f$, where $r$ is the number of temporal basis functions in \eqref{bi-linear}. 

\section{Experiments and Results}
We describe the experiments and our observations in this section.
\subsection{Impact of pretraining}
\label{pretrain}
To demonstrate the benefit of pretraining, we compare the algorithms described in \eqref{DEBLUR}  with random initialization and with the initialization scheme denoted in \eqref{thetainit} \& \eqref{phiinit}. Fig \ref{fig2}.(a) shows the plot of SER values with respect to the number of epochs. The color of the borders of the images in Fig \ref{fig2}.(b)-(e) correspond to the color of the markers in the plots, which denotes the initial and maximum SER values. We note that the initialization with random weights starts with the low SER as seen from Fig. \ref{fig2}.(b), compared to the SToRM initialization in Fig \ref{fig2}.(d). As seen from the plots in (a), the DEBLUR approach with STORM initialization converged rapidly to a peak value, while the one with random initialization converged to a poor solution with significantly more iterations. We also note that the DEBLUR approach with random initialization yielded poor results as seen from Fig. \ref{fig2}.(c), while the one with STORM initialization in \ref{fig2}.(f) shows significantly improved results. 

\begin{table}
	\caption{Performance comparison of the DEBLUR method using the BRISQUE score on multiple data. We have also shown the benefit of using more acquired data. Corresponding images are shown in Fig \ref{fig6}}
	\centering
	\setlength{\tabcolsep}{3pt}
	\begin{tabular}{|p{45pt}|p{35pt}|p{35pt}|p{35pt}|p{35pt}|p{35pt}|}
	    \hline
		Method & Data1 & Data2 & Data3 & Data4 & Data5\\
		\hline
		DEBLUR(42s) & 
		$28.30\pm3.7$ & 
		$30.33\pm3.3$ &
		$29.90\pm3.8$ &
		$28.11\pm3.7$ &
		$17.53\pm5.0$\\
		\hline
		SToRM(42s) & 
		$34.44\pm3.9$ & 
		$26.0\pm2.9$ &
		$29.96\pm4.3$ &
		$34.0\pm5.0$ &
		$17.32\pm5.4$ \\
		\hline
		DEBLUR(14s) & 
		$29.5\pm5.3$ & 
		$25.94\pm3.4$ &
		$28.61\pm4.1$ &
		$32.87\pm5.5$ &
		$17.28\pm5.2$ \\
		\hline
		SToRM(14s) & 
		$36.26\pm0.9$ & 
		$32.63\pm1.3$ &
		$35.72\pm3.5$ &
		$34.57\pm0.8$ &
		$25.42\pm2.3$\\
		\hline
	\end{tabular}
	\label{tab3}
\end{table}

sub\subsection{Impact of regularization parameters}
\label{regpar}
We study the impact of the regularization priors in \eqref{DEBLURrg} in Figures \ref{fig3}, \ref{fig4} and \ref{fig4a}.  

\subsubsection{Regularization of $\mathbf U$ network parameters}
We initialize the network with the SToRM initialization, described by \eqref{unet} \&  \eqref{vnet}. We set $\lambda_2=\lambda_3=0$ and consider four different $\lambda_1$ values in this study. We note that the network with $\lambda_1=0$ results in the performance peaking after a few iterations, similar to the case in Fig. \ref{fig2}. With more iterations, the SER drops because of the overfitting to noise. By contrast, we observe that $\lambda_1=10^{-3}$ results in saturating performance, which is around 2.5 dB superior to the peak performance obtained from $\lambda_1=0$. We also observe that $\lambda_1$ values that are slightly higher and lower than the optimal values result in somewhat similar performance, which indicates that the network is not very sensitive on the optimal values.

\subsubsection{Regularization of $\mathbf V$ network parameters}
In Fig. \ref{fig4}, we study the impact of $\lambda_2$ on the results. We keep the best  $\lambda_1=10^{-3}$ value from Fig. \ref{fig3} and set $\lambda_3=0$. We vary $\lambda_2$ in this study and plot the change in SER with iterations. We note that $\lambda_2=10^{-4}$ offers the best final performance, resulting in around 0.2 dB improvement in performance compared to $\lambda_2=0$. We note that the joint optimization of $\mathbf V$ and $\mathbf U$ networks results in around 8-9 dB improvement in performance over the SToRM initialization. The networks learned by DEBLUR result in a bilinear representation that is more optimal in representing the data when compared to the classical bilinear methods. 

\subsubsection{Regularization of latent vectors $\mathbf Z$}
In Fig. \ref{fig4a} we study the impact of $\lambda_3$ on the results. We keep $\lambda_1=10^{-3}$ and $\lambda_2=10^{-4}$, which were the best values that we determined in the previous subsections. We consider different values of $\lambda_3$ and plot the corresponding latent vectors. We observe that the latent vector regularization had marginal impact on the SER. We note that the optimal value is $\lambda_3=1e^2$. We observe that $\lambda_3=0$
resulted in noisy latent vectors, while the one with $\lambda_3=1e^2$ offered the disentangling of cardiac/respiratory motion.

\subsubsection{Benefits of using latent vectors to generate temporal basis}

We note from Fig \ref{fig4a} and Fig. \ref{fig5} that the $\lambda_3$ parameter is not very sensitive to image quality. However, by properly selecting $\lambda_3$, we observe that the latent vectors learn physiologically relevant parameters. In the datasets we considered, we note that the latent vectors separate into a fast-changing latent vector that captures cardiac motion and a slow one that captures respiratory motion. Post-reconstruction, the latent variables can be used to bin the images into different cardiac and respiratory states. The top row corresponds to the latent vectors and reconstructions from 14 s of data. The red box (corresponding to the red lines in the latent vector plot in (a)) corresponds to the image frame in the systole phase and expiration state, while the blue box corresponds to an image in the diastole phase and expiration phase. Fig \ref{fig5}(c) shows the latent vectors estimated from 28 s of data. The results show that similar decomposition of latent vectors can be obtained from more data, with moderate improvement in image quality.  

\subsection{Comparison with state-of-the-art methods}
We compare the proposed DEBLUR method with the SToRM and low-rank reconstructions. Fig \ref{fig6} shows the comparison between the DEBLUR and SToRM methods using five different datasets. We consider the recovery from 42 s and 14 s of data. We observe that the DEBLUR(42s) visually offers similar or improved image quality when compared to SToRM(42s), manifested by reduced blurring and the depiction of the papillary muscles. 

The SToRM approach results in significant blurring and degradation in image quality when only 14 s of data is available. We note that the DEBLUR reconstructions are able to minimize the blurring, offering reconstructions that are comparable to the 42 s reconstructions. The improved performance may be attributed to the spatial and temporal regularization of the factors offered by DEBLUR. 

Since ground truth is not available, we have used SToRM(42s) acquisition as reference data for SER and SSIM comparisons in Table \ref{tab2}. We also compare the methods using the BRISQUE score in Table \ref{tab2} and \ref{tab3}, where we show the BRISQUE scores of the individual datasets. 

In Fig \ref{fig7}, we compare the methods that recover the images from 14 s of data against SToRM reconstructions from 42 s. We have shown two frames (end of diastole and end of systole) from each method, along with error maps in Fig \ref{fig7}(a). Time profiles are depicted in Fig \ref{fig7}(b). We note that DEBLUR provides the best spatial and temporal quality and improved details, which are comparable to the SToRM reconstructions from 42 s. 

\section{Discussion}
The experiment in Section \ref{pretrain} clearly shows the benefit of initializing the network parameters by \eqref{thetainit} and \eqref{phiinit}, respectively. As seen from Fig. \ref{fig2}, the optimization process is able to offer around 8-9 dB improvement in performance over the SToRM initialization. We note that the proposed framework of recovering the bilinear representation from 10 spokes/frame is significantly more challenging than the traditional DIP strategy in \cite{dip}. A reasonable initialization of the network can offer improved performance compared to random initialization. Fig. \ref{fig2} also shows the need for early stopping of the unregularized setting in \eqref{DEBLUR}. In particular, the performance of the algorithm decays with increasing iterations, indicating overfitting to the noise in the k-space measurements. 

These experiments in Section \ref{regpar} show the benefit of regularizing the generator parameters while fitting to undersampled data. We observe from Fig. \ref{fig3} \& \ref{fig4} that regularizing the network parameters improves the generalization performance of the network to unobserved k-space samples. Specifically, the network parameters are learned from few measured k-space data. The regularization of these networks reduces the impact of overfitting, thus improving the degradation in performance with iterations. 

The best choice of $\lambda_3$ enables the learning of latent vectors $\mathbf Z$ that are interpretable. In particular, the latent vectors are observed to learn the temporal variations seen in the data, including cardiac and respiratory motion. Interpretable latent vectors can aid in visualization of the results, as showcased in Fig. \ref{fig5}. In particular, the slow-changing latent vector captures the respiratory motion, while the fast latent vector captures the cardiac motion. The data can be sorted into the respective phases based on the latent vectors.

The comparison of the proposed scheme with the classical approaches shows that the proposed DEBLUR(42s) can offer comparable performance to  SToRM(42s), while the proposed scheme from 14 s of data significantly outperforms the SToRM(14s) approach.

\section{Conclusion}

We introduced a deep unsupervised bilinear algorithm to reconstruct dynamic MRI from undersampled measurements. We represented the spatial and temporal factors using two CNN- based generators, which are learned from the undersampled k-space data of each subject. The initialization of the network weights using an existing bilinear model (SToRM) is observed to both reduce the run-time as well as offer improved performance compared to initialization by random weights. The weights of the networks are regularized with $l_1$ penalty to minimize the overfitting of the network to the noise in the measurements. The weight regularization is observed to minimize the degradation in performance with iterations. The implicit and learned regularization offered by the proposed scheme offers improved image quality compared to current methods, especially while recovering the data from shorter acquisition strategies. The proposed scheme directly learns a compressed image representation from the measured data, making it considerably more memory efficient than current approaches. The high memory efficiency makes it readily applicable to large-scale dynamic imaging (e.g., 3D+time) applications. In addition, the unsupervised strategy also eliminates the need for fully sampled training data, which is often not available in large-scale imaging problems. 

\section{COMPLIANCE WITH ETHICAL STANDARDS}
This research study was conducted using human subject data. The Institutional Review Board at the local institution approved the acquisition of the data, and written consent was obtained from the subject.
\section{ACKNOWLEDGMENTS}
This work is supported by grant NIH 1R01EB019961. The authors claim that there is no conflict of interest.
% \section{Conclusion}
% \label{sec:con}
% In this paper, we have proposed a new cardiac cine MRI reconstruction method based on bilinear unsupervised learning. We have used regularized spatial and temporal priors to utilize the benefits of a bilinear model in dynamic image reconstruction and to avoid early stopping constraint.  Results show that our CNN priors with $l_1$ norm on the learning parameters give better performance when compared to the deep image prior. Reconstructed cardiac cine images show the ability of our proposed method to yield improved image quality over other methods.

\bibliographystyle{IEEEtran}
\bibliography{refs_tmi}

\end{document}